\documentclass[12pt]{article}

\usepackage{mathtools} 
\mathtoolsset{showonlyrefs}

\usepackage{amsmath, amsfonts,amssymb,hyperref}
\usepackage{url}
\usepackage{graphicx}%
\usepackage{bbm}

\makeatletter

\@addtoreset{equation}{section}
\def\section{\@startsection {section}{1}{\z@}{-2.5ex plus -1ex minus
 -.2ex}{1.3ex plus .2ex}{\large\bf}}
\def\subsection{\@startsection{subsection}{2}{\z@}{-2.25ex plus%
 -1ex minus -.2ex}{0.5ex plus .2ex}{\bf}}

\advance \voffset by -1.0in
\advance \hoffset by -0.6in
\def\bpm{\begin{pmatrix}}
\def\epm{\end{pmatrix}}

\newcommand{\NN}{\mathbb{N}}
\newcommand{\ZZ}{\mathbb{Z}}

\newcommand{\RR}{\mathbb{R}}

\def\bee{\begin{equation}}
\def\eee{\end{equation}}
\newtheorem{theorem}{Theorem}[section]

\newtheorem{lemma}[theorem]{Lemma}

\begin{document}

\vskip 10pt
\baselineskip 28pt

\begin{center}
{\Large \bf Spectral Geometry of Nuts and Bolts }

\baselineskip 18pt

\vspace{0.4 cm}

{\bf  Lyonell Boulton, Bernd J~Schroers and  Kim Smedley-Williams}\\
\vspace{0.2 cm}
Maxwell Institute for Mathematical Sciences and
Department of Mathematics,
\\Heriot-Watt University,
Edinburgh EH14 4AS, UK. \\
{\tt l.boulton@hw.ac.uk}, \;\;  {\tt b.j.schroers@hw.ac.uk}  and {\tt ks275@hw.ac.uk}

\vspace{0.4cm}

{May  2022 }

\end{center}

\begin{abstract}
\noindent  We study the spectrum of  Laplace operators on a one-parameter family of  gravitational instantons of bi-axial Bianchi IX type coupled to an abelian connection with self-dual curvature. The  family of geometries includes the Taub-NUT, Taub-bolt and Euclidean Schwarzschild geometries and interpolates between them. The interpolating geometries have conical singularities along a submanifold of co-dimension two, but we prove that the associated Laplace operators have natural selfadjoint extensions and study their spectra. In particular, we determine the essential spectrum and prove that its complement, the discrete spectrum, is infinite. We compute some of these eigenvalues numerically and compare the numerical results with an analytical approximation derived  from the asymptotic Taub-NUT form of each of the geometries in our family. 
\end{abstract}

\baselineskip 14pt
\parskip 4 pt
\parindent 10pt

\baselineskip 16pt

\section{Introduction}

Laplacians on manifolds with non-trivial Riemannian metrics arise  in rather diverse contexts in mathematical physics.  The simplest and most familiar example is the Laplacian on the Lie group $SO(3)$ equipped with a left-invariant metric, whose spectrum describes the quantum excitations of a rigid body. More involved examples arise generally when collective coordinates are used to model a physical system  where kinetic effects   are important. Mathematically, such effects are captured by an induced Riemannian metric on the space of collective coordinates, and in the corresponding quantum theory the associated Laplace operator is a natural candidate for the  kinetic part of the Hamiltonian. This point of view is systematically explored in the study of collective coordinates for solitons, often called moduli spaces in this context \cite{AHbook,GM}. 

The spectrum of Laplacians associated to four-dimensional Riemannian geometries  which satisfy the Einstein equations has received particular attention in the literature. Such spaces were first studied  in the  context of the path-integral approach to quantum gravity and are called gravitational instantons \cite{Hawking_instantons}\footnote{In differential geometry the term gravitational instanton is often reserved for the more restricted class of  hyperk\"ahler metrics, but we will use the original, wider definition here.} . 
The spectrum of the Laplace operator associated to a gravitational instanton is important in understanding the fluctuations around the instanton, and in determining its semiclassical contribution to the path integral.   Gravitational instantons also  occur as  moduli spaces of magnetic monopole \cite{AHbook}, and here the Laplace operator can be used to study quantum properties of monopoles \cite{GM}.

It is often natural to couple the Laplace operator to a connection on the Riemannian manifold. In the context of collective coordinates, such connections may be induced as Berry connections when additional linear collective coordinates are included in the model. In the context of gravitational instantons, the relevant connections are often themselves  Yang-Mills instantons,  i.e., their curvatures obey equations which are mathematically natural on four-dimensional Riemannian manifolds. 

In this paper we examine the spectral properties of a family of non-compact four-dimensional Riemannian geometries which  obey  the Einstein equations and which permit isometric actions of $(SU(2)\times U(1))/\ZZ_2$. The family includes three much studied spaces - the self-dual Taub-NUT geometry (TN), the Taub-bolt space (TB) and the Euclidean Schwarzschild space (ES) - but also interpolating geometries which have so-called edge-cone singularities \cite{ALB}. Each geometry comes with an abelian connection whose curvature 2-form is self-dual (an abelian instanton), and in each case we consider the Laplace operator minimally coupled to that connection.   

The  essential difference between  TN, TB and ES lies in the geometry  of the  $U(1)$ orbits. Each space has an  interior region  where the $U(1)$-action has fixed points. There is a single fixed point, called a nut, for TN, but a 2-sphere of fixed points, called a bolt,  for ES and TB \cite{GH}. Away from the fixed points, the spaces are  principal $U(1)$ bundles over a three-dimensional space which is conformal to  flat Euclidean space. The radius of the $U(1)$-orbits tends to a constant value  as the proper distance from the fixed points tends to infinity.  At a fixed positive distance, the orbits are fibres of a non-trivial fibre bundle over a 2-sphere in the case of TN and TB, but fibres of a trivial fibration for ES.

Our motivation for studying the spectral properties of this family of geometries is two-fold.  The family interpolates between the  TN and ES geometries  whose spectral properties were studied in \cite{JS2} and \cite{JS3}, with several interesting results. The  spectrum of TB has not been studied, so our first motivation is  to fill that gap and to provide a unified understanding of the spectra associated to  three superficially rather different geometries. Also, with a bolt in its interior and a non-trivial bundle structure asymptotically,  TB is the most interesting of the three spaces. 

Our second motivation stems from the conical singularities in the interpolating geometries. Geometries with conical singularities on submanifolds of real co-dimension two  are of interest in differential geometry and analysis, see \cite{ALB} and references therein. In particular, there are two famous one-parameter families of such edge-cone geometries, one due to  Hitchin \cite{Hitchin} and one due to Abreu \cite{Abreu}, which involve spaces closely related to the ones we study here. Unlike our manifolds, the manifolds considered by Abreu and Hitchin are compact except for limiting cases, which are, respectively, the Taub-NUT and Atiyah-Hitchin spaces.  Edge-cone singularities may be problematic for the classical geometry, but here we show that, at least in our examples, they do not pose any difficulties for the spectral theory. The  Laplacians  we consider have natural selfadjoint extensions and infinitely many bound states even when the corresponding classical geometry has conical singularities.  This echoes observations about Laplacians on two-dimensional cones in \cite{KS}.

The smoothing out of the conical singularities in the spectral theory is reminiscent of a similar effect for singular potentials in standard Hamiltonians studied in atomic and nuclear physics, where the quantum version (the spectral theory) is well-behaved, even when the classical theory (trajectories obeying suitable differential equations) is not.  Since Laplacians associated to Riemannian geometries are increasingly used to model quantum effects, it is natural to explore the robustness of these operators when
 singularities are introduced in the classical picture.

Our paper is organised as follows. In Section 2 we discuss the geometry of  gravitational instantons  of the so-called Bianchi IX form which include and   interpolate between the TN, TB and ES geometries, paying particular attention to the  interpolating geometries with conical singularities. We also introduce a family of self-dual 2-forms on these geometries, and interpret them as curvatures of a family of $U(1)$ connections. In Section 3  we define the Laplace operators on the family of geometries,  minimally  coupled to the family of $U(1)$ connections.  Exploiting the symmetry and paying careful attention to the topology of the $U(1)$ bundle over the bolt for the TB and ES spaces, we separate variables and reduce the study of gauged Laplace operators to the study of radial Hamiltonians, defined on the half-line.  The selfadjointness and spectral properties of the radial Hamiltonians are the subject of Section 4, which contains our main analytical results. The radial Hamiltonians depend on geometric parameters and on parameters characterising  angular momenta of the functions used in separating variables.  The intricate dependence of spectral properties on these two sets of parameters, which one may call classical and quantum, and on their interplay is the focus of this section.  The TN geometry is special among the geometries studied here  in that both its geodesics and its spectrum can be determined analytically, even when the coupling to the self-dual connection is included \cite{JS2}. In Section 5 we exploit the fact that all  Hamiltonians studied in this paper have an asymptotic form of the TN type to compute analytical approximations to the discrete spectrum  of the TB family of metrics. These analytical results are then compared with a numerical study of the discrete spectrum for illustrative choices of the parameters, using a Riesz-Galerkin approximation. We end the paper with a  brief discussion of the results, and with our conclusions.

\section{Geometry of nuts and bolts in  bi-axial Bianchi IX geometries}
\subsection{A family of Taub-bolt spaces and its limits}
The gravitational instantons we want to consider have sufficient symmetry to reduce both the Einstein equations and  the eigenvalue problem for the  Laplace operator to problems involving ordinary differential equations. Technically, this means that we consider four dimensional Riemannian manifolds with  isometric   group actions whose orbits  are generically three dimensional, i.e.  group actions of co-homogeneity one.  If the isometry group is $SU(2)$ and the action has generically three-dimensional orbits, the Einstein condition implies  that the metric can be expressed in the   Bianchi IX form \cite{DS}
\bee
\label{genbianchi}
ds^2 = f^2 dr^2 + a^2\sigma_1^2 + b^2\sigma_2^2 + c^2 \sigma_3^2, 
\eee
where $\sigma_1,\sigma_2$ and $\sigma_3$ are left-invariant 1-forms on $SU(2)$ and $a,b,c$ and $f$ functions of a transverse or radial coordinate which will denote as $r$ in the following. 

We adopt the conventions of  \cite{JS2} for our parametrisation of $SU(2)$ in terms of Euler angles, but name the angles $(\phi,\theta,\psi)$ instead of the $(\alpha,\beta,\gamma)$  used in \cite{JS2}. With the ranges $0\leq\phi < 2\pi,\;  0\leq \theta \leq  \pi, \; 0\leq \psi <4\pi$,  an element $g\in SU(2)$ is parametrised in terms of Lie algebra generators $t_i, i=1,2,3$,  with commutators $[t_i,t_j]=\epsilon_{ijk}t_k$  as 
\bee
g = \exp(\phi t_3) \exp(\theta t_2)\exp(\psi t_3).
\eee
These coordinates need to be used with care when $\theta=0$ or $\theta=\pi$,  where neither $\phi$ nor $\psi$ is  individually well-defined. However, for us it will be important that  $\phi+\psi$ is well-defined when $\theta=0$ and $\phi-\psi$ is well-defined when $\theta=\pi$.
Written in terms of these angles, the left-invariant 1-forms  on $SU(2)$  are 
\begin{eqnarray}\label{LI}
 \sigma_1 &=&  \sin\psi d\theta - \cos\psi \sin\theta d\phi,  \nonumber\\
 \sigma_2 &=& \cos\psi d\theta + \sin\psi \sin\theta d\phi, \nonumber\\
\sigma_3 &=&  d\psi  + \cos\theta d\phi.
\end{eqnarray}
They satisfy $d\sigma_3 = -\sigma_1\wedge\sigma_2$ and its cyclic permutations. 
For our study of the Laplace-Beltrami operator we also require the basis  of vector fields  dual to left-invariant 1-forms, i.e. satisfying $\sigma_i(\xi_j) =\delta_{ij}$. In terms the Euler angles, they are 
\begin{align}\label{dual}
 \xi_1 &= \phantom{-}\cot\theta \cos \psi \partial_\psi
           + \sin \psi \partial_\theta
            -\frac {\cos \psi }{ \sin\theta} \partial_\phi,
\nonumber\\
\xi_2 &= -\cot\theta \sin \psi \partial_\psi
           + \cos \psi \partial_\theta
            + \frac {\sin \psi}{ \sin\theta} \partial_\phi,
\nonumber \\
 \xi_3 &= \phantom{-}\partial_\psi.
\end{align}

The function $f$ can be chosen to fix the radial coordinate, but $a,b$ and $c$  satisfy a system of second order ordinary differential equations by virtue of the Einstein condition. In general it is a difficult problem to determine and classify all solutions of these equations which result in smooth Riemannian geometries. However, with the assumption of an additional $U(1)$ symmetry, this problem simplifies and is easily solved. This is the situation we study in the present paper. The additional symmetry requires that $a=b$, leading to the so-called bi-axial Bianchi IX  type, and the Einstein equations are equivalent to a system of coupled  ordinary differential equations \cite{GP,AD}.  Writing a dot for the derivative $\frac 1 f \frac{d}{dr}$ with respect to proper radial distance,  the system consists 
of  two second order evolution  equations
\bee
\label{evolution}
  \frac{\ddot{a} }{a}-  \frac{\dot{a}}{a}\frac{\dot{c}}{c} + \frac{c^2}{4a^4} = 0, \quad 
   \frac{\ddot{c} }{c} + 2 \frac{\dot{a}}{a}\frac{\dot{c}}{c}- \frac{c^2}{2a^4}  =0,
\eee
and the conservation law
\bee
\label{constraint}
 2 \frac{\dot{a}}{a}\frac{\dot{c}}{c} +  \left(\frac{\dot{a}}{a}\right)^2 -\frac{1}{a^2}+\frac{c^2}{4a^4}=0.
\eee
The general solution  of this system depends on three constants, one of which can be viewed as an overall additive shift in the proper radial distance. The most widely used form of the metric   is based on the choice  $f= 2N/c$ and  depends on two  real parameters $N$ and $M$ which can be chosen to be positive without loss of generality \cite{Page} and which  are ordered  $M>N$ for a complete metric without boundary \cite{GP}. The solution is conveniently expressed in terms of  the function
\bee
\label{Upot}
U =  \frac{r^2- 2Mr + N^2} {r^2-N^2},
\eee
and takes the form 
\bee
a=b=\sqrt{r^2-N^2}, \quad c=2N \sqrt{U}, \quad f=\frac{1}{\sqrt{U}},
\eee
so that 
\bee
\label{genmet}
ds^2 = \frac 1 U dr^2 + (r^2-N^2) (\sigma_1^2 + \sigma_2^2) +4N^2 U \sigma_3^2.
 \eee
 This is the general family of  metrics with which we work in this paper. We will use the orientation determined by the volume form
\bee
\label{volume}
dV = 2N(r^2-N^2)   \sigma_1\wedge \sigma_2\wedge \sigma_3\wedge dr.
\eee
Factorising 
\bee
 r^2- 2Mr + N^2 =(r-r_+)(r-r_-), \qquad  r_\pm = M\pm\sqrt{M^2 -N^2},
\eee
and defining a  parameter $k\geq 1$ via 
\bee
r_+ = kN, \quad r_-=\frac N k, 
\eee
it is necessary (but not sufficient)  to take the range of  $r$ as   $[kN,\infty]$  for a complete and non-singular metric.   This range avoids the  curvature singularities at the poles $r=\pm N$ of $U$ \cite{Page}, and  the zero of $U$ at $r=kN$  ensures that this is not a boundary of the manifold but, in the terminology introduced in \cite{GH},  either a nut or a bolt. The former occurs when $k=1$ and means that the $SU(2)$ orbit collapses to a point. The latter occurs when $k>1$ and means  that the $SU(2)$ orbit collapses to  the  2-sphere parametrised by the  polar angles $(\theta,\phi)$.  In this paper we think of the metrics  \eqref{genmet}   as a two-parameter family, parametrised by $N>0$ and $k\geq 1$.  We  are particularly interested in the dependence on $k$ and only comment on the role of $N$ briefly in our conclusions. 

To avoid   conical singularities at the bolt one  in fact requires $k=2$, as can be seen by switching to the coordinate $\varrho=\sqrt{r-r_+}$.
Near the bolt (at $r=r_+$) one checks that  $U\approx \varrho^2/(kN)$ so that, keeping leading terms in $\rho$, the metric is approximately
\bee
ds^2 \approx 4Nk \left[(d\varrho)^2 +  \varrho^2\frac {\sigma_3^2}{k^2}\right] + (k^2-1)N^2 (\sigma_1^2+\sigma_2^2).
\eee
This is a metric on  the  branched cover of cotangent bundle to the 2-sphere. The metric on the two-dimensional  cotangent plane at  fixed   $\theta$ and $\phi$  is approximately
\bee
ds^2  \approx 4Nk \left[ (d\varrho)^2 +  \varrho^2\frac {d\psi ^2}{k^2}\right] .
\eee
With $\psi$ in the interval $[0,4\pi]$,  this is the flat metric on the plane precisely if  $k=2$ but when $k\neq 2$ it is the geometry of a singular cone.  The smooth geometry obtained when $k=2$, first discovered by Page in \cite{Page}, corresponds to TB.   In this case, $r_+=2N,  r_-=\frac 1 2 N, M=\frac 5 4 N$ and so 
\bee
U =\frac{r^2-\frac 5 2 Nr + N^2} {r^2-N^2}.
\eee

The case $k=1$ has to be treated separately since $r_+=r_-=N=M$ in this case, leading to a cancellation and reduction in the order of the rational function  $U$, which is now $U=\frac{r-N}{r+N}$. 
 The resulting space is the Euclidean Taub-NUT gravitational instanton \cite{Hawking_instantons},  so TN in our notation.  Its metric 
 \bee
\label{TNmetric}
ds^2= \frac{r+N}{r-N} dr^2 + (r^2-N^2)(\sigma_1^2 +\sigma_2^2) + 4N^2
\frac{r-N} {r+N} \sigma_3^2, \qquad r \geq N,
\eee
is frequently expressed in terms of the alternative radial parameter $  \rho = r-N$. Then, with $\Lambda=2N$, 
 it takes the isotropic form 
\bee
\label{TNisotropic}
ds^2= \left(1+\frac{\Lambda}{\rho}\right) \left( d\rho^2 + \rho^2 (\sigma_1^2 +\sigma_2^2)\right) + \Lambda^2
\frac{\rho} {\rho+\Lambda } \sigma_3^2,\quad \rho \geq 0, 
\eee
 used  in \cite{JS1}\footnote{The constant $L$ in that paper is denoted $\Lambda$ here to avoid a notational clash later in our discussion.}. The apparent singularity at $r=N$ or $\rho=0$ is only a coordinate singularity. It can be removed by working with $\varrho= \sqrt{r-N}$. Keeping leading terms only, one finds for small $\varrho$ that
 \bee
 ds^2\approx 8N\left[d\varrho^2 + \frac {\varrho^2}{4} (\sigma_1^2 + \sigma_2^2 + \sigma_3^2)\right].
 \eee
 This  shows that the geometry near $r=N$ is  that of  Euclidean $\RR^4$, with the nut being the origin.

Even though the classical geometry is only smooth for $k=1$ or $k=2$, we will treat $k$ as a real parameter in the range $k\in [1,\infty)$ in this paper. This means that we allow conical singularities at the bolt. We refer to the metrics with $1<k<\infty$ as the TB family, and sometimes simply as TB spaces.

  Relative to the flat angular range $2\pi$, the cones have a surplus angle for $1<k<2$ and deficit angle  for $k>2$ given by  
\bee
\Delta \psi  =\left(\frac {2-k}{ k}\right)2\pi.
\eee
It is remarkable that in the limit $k\rightarrow 1$,  as the surplus angle approaches $2\pi$ and the bolt collapses to a nut, the  Riemannian geometry near the bolt  jumps from that of the cotangent bundle of $S^2$ to that of $\RR^4$.

The discussion so far assumes that the range of the angle $\psi$ is $[0,4\pi)$. However, it is geometrically natural to consider  Bianchi IX manifolds whose generic $SU(2)$ orbit is the Lens space $S^3/\ZZ_n$, where the generator of $\ZZ_n$ acts by sending $\psi \mapsto \psi +\frac{4\pi}{n}$  for some positive integer $n$.
In our parametrisation this simply leads to the shorter range  $\psi \in \left[0,\frac{4\pi}{n}\right)$.  Repeating the above analysis with this range,  the requirement of a smooth geometry at the bolt  when $k>1$ now requires $kn=2$, which is only possible if  $n=1$ and again $k=2$, while in the case $k=1$  smoothness immediately yields $n=1$ \cite{Page}. However, if one permits conical singularities as we do here, we can allow any positive integer $n$ in the case where  $k>1$, i.e. when  there is a bolt at $r=r_+$. The conical geometry always has a deficit angle when $n\geq 2 $, with deficit angle $  4\pi /(kn) -2\pi$.
 
 All the geometries we have considered so far are asympotically locally flat, or ALF in the terminology used for classifying gravitational instantons. In our examples this means that for large $r$ (but in fact  $r>r_+$ is sufficient), the spaces are topologically circle bundles over $\RR^3\setminus \{0\}$, with the  circle (parametrised by $\psi$ here)  having a finite radius which tends to  $4N/n$ for $r\rightarrow \infty$. At fixed $r>r_+$ they are the non-trivial circle bundles over $S^2$ whose total space is the Lens space $S^3/\ZZ_n$.

To end this review of the classical geometry, we observe that two further rotationally symmetric gravitational instantons  with bolts can be obtained from the general solution \eqref{genmet} by taking limits, namely the Eguchi-Hanson space and the Euclidean Schwarzschild space.  The details are summarised in \cite{Page}    but we note that  the Eguchi-Hanson space is obtained 
for $n=2$  by taking the limit  $ k\rightarrow 1$  while keeping $ 4(k^2-1)N^2$ constant (so  that necessarily  $N\rightarrow  \infty$). 
It  has a self-dual  Riemann curvature and is an example of an asymptotically locally Euclidean (ALE) space:  for fixed radial distance from the bolt, the geometry is that of a non-trivial circle bundle over $S^2$ and for large radial distances the radius of both  the circle and the sphere grow linearly with that distance.

The ES  space is obtained  by  picking  a  positive constant  $L$,   and taking  the limit
\bee
\label{geolim}
N=\frac{L}{k},\quad k\rightarrow \infty .
\eee
This implies that $r_+=L$, but  $r_-\rightarrow 0$.
We also change angular coordinates according to 
\bee
\chi = \frac{\psi}{k}= \frac{N}{L}\psi,
\eee
and  observe that in the ES limit
\bee
2N\sigma_3= 2N \cos\theta d\phi + 2 L d\chi \rightarrow 2 L d\chi.
\eee 
Then
\bee
U \rightarrow  V= 1-\frac{L}{r}, \qquad 4N^2\sigma_3^2  \rightarrow  4L^2d\chi^2,
\eee
and the metric becomes
\bee
\label{ESmetric}
ds^2= \frac{1}{V}dr^2 + r^2(\sigma_1^2 + \sigma_2^2) +4L^2  Vd\chi^2,\quad r\geq L, 
\eee
which is the standard form  of the ES metric, as studied in  \cite{JS2}.
For a fixed positive distance from the bolt, the topology is that  of  a trivial circle bundle over $S^2$.

\subsection{Magnetic fields}

The study of harmonic 2-forms on Riemannian 4-manifolds is interesting for a number of apparently different reasons. Physically, such forms represent solutions of the Maxwell equations, albeit in a Riemannian rather than the Lorentzian setting \cite{Pope}.  Geometrically, they  arise as the curvature of abelian connections on  the manifold.  In differential topology, the existence of harmonic 2-forms is related to the homology of the 4-manifold.  The situation is simplest in the compact setting, where harmonic forms are the Hodge representatives of cohomology classes. In the non-compact setting, the relationship between harmonic forms and homology of the underlying manifold depends on asymptotic behaviour, but happily  this has been studied in detail for  the TB and TN manifolds recently \cite{Franchetti}.
The final reason why harmonic forms are interesting is particularly relevant here. Coupling linear differential operators naturally associated to a Riemannian manifold - such as Dirac or Laplace operators - to abelian connections enriches the associated spectral theory. For example,  on the TN space, the Laplace operator has no discrete spectrum, but the same operator  coupled  minimally to an abelian connection  with self-dual and square-integrable curvature has infinitely many bound states \cite{JS2}. 

Here we are interested in a family of connections of the TN-TB family of metrics \eqref{genmet} with $k\geq 1$ which generalises the self-dual connection on the TN space. As explained in \cite{Franchetti}, the TB space $(k=2)$ has a two-dimensional space of harmonic 2-forms, which contains a one-dimensional subspace of self-dual 2-forms. These forms can easily be generalised for any value  $k>1$ in such a way that they become a self-dual form on TN in the limit $k\rightarrow 1$. 
 The resulting family of forms, self-dual with respect to the volume element \eqref{volume},  can be written as 
 \bee
 F= \frac p 2\frac{k+1}{k-1} \left( \frac{2N}{(r+N)^2} dr  \wedge \sigma_3 + \frac{r-N}{r+N}\sigma_2\wedge\sigma_1\right).
 \eee
 Our choice of normalisation for $F$  and the interpretation of the parameter $p$ follow  from our wish to treat  $iF$ as the curvature of a connection on the TB manifold. At the bolt, 
  \bee
  iF|_{r=kN}= -i\frac p2 \sin\theta d\theta\wedge d\phi,
  \eee
 showing that our normalisation implies that,  for $k>1$,  $p$ is the first Chern class of the connection $iA$ over the bolt:
  \bee
  \label{chern}
  \frac{i}{2\pi} \int_{\text{\tiny bolt}} (iF)= p.
  \eee
  The integrality of the first Chern number is therefore equivalent to 
 \bee
 \label{pnorm}
  p\in \ZZ.
 \eee

 The  curvature form  is not globally exact, but away from the bolt, $iF=idA$ with 
 \bee
 \label{genA}
 iA= i\frac p 2\frac{k+1}{k-1}\frac{r-N}{r+N}\sigma_3.
 \eee 
 This formula fails at the bolt since  the angle $ \psi$ and hence $\sigma_3$ is not well-defined there. 
   We will use $iA$ as a gauge potential  in our gauged version of  the Laplace operator, but to obtain a connection over TB we also need gauge potentials on the bolt. They can be constructed in analogy with the    gauge potentials used in the discussion of the Dirac monopole. Defining local gauge transformation $g_\pm=\exp(i\frac p 2 (\psi\mp \phi))$, and monopole gauge potentials via 
 \bee
 \label{gaugepot}
 iA_\pm = iA -  g_\pm^{-1}dg_\pm ,
 \eee
 one checks that,  at the bolt, 
  \bee
 \label{mopoA}
  iA_\pm|_{r=kN}=i \frac p 2 (\cos\theta \pm 1) d\phi,
 \eee
 so that $A_+$ is  regular in the  southern ($\theta >0$) and $A_-$ in the northern ($\theta <\pi$)  coordinate patches covering the bolt. 
 In practice,   one can work in the smooth monopole gauge, with two gauge potentials $iA_+$ and $iA_-$, smooth in the open sets $\theta >0$ and $\theta <\pi$ and related on their overlap by the smooth gauge transformation $g=\exp(i p  \phi)$, or one can apply the gauge transformation $g_\pm$, singular at the bolt, to  work with the gauge potential $iA$, also singular at the bolt.

 The  necessity of local gauge potentials, related by local gauge transformations, suggests that in studying the spectrum of the gauged Laplace operator one may have to work with local sections, also related by gauge transformations. Explicitly, writing $\Phi$ for a section in the singular gauge, and $\Phi_\pm$ for local sections in the monopole gauges, we need to combine \eqref{gaugepot} with 
 \bee
 \label{Phigauge}
 \Phi_\pm = g_\pm\Phi
 \eee
 to satisfy the covariance condition $(d+iA_\pm)\Phi_\pm =g_\pm (d+iA)\Phi$.  Happily, this is not necessary in practice. As in the analogous treatment of the  gauged Laplace operator on the ES space in \cite{JS3},  one can  exploit the covariance of the gauged Laplace operator to work with the  singular gauge potential $iA$, even at the bolt, but then determine the allowed solutions $\Phi$ by requiring that  the gauge transformed sections \eqref{Phigauge} are smooth at the bolt. We will see in the next section that this requirement can be implemented very naturally in practice.

A gauge potential for the TN space can formally be  obtained by taking the limit $k\rightarrow 1$ and simultaneously re-scaling $A$ by taking $p\rightarrow 0$ in such a way that $\frac{k+1}{k-1}p\rightarrow  \tilde p$ for some $\tilde p \in \RR$.  The resulting 1-form is\footnote{The parameter $\tilde p$ is denoted $p$ in \cite{JS1,JS2}.} 
\bee
\label{TNswitch}
A_{\text{\tiny TN} }= \frac{\tilde p}{2}\frac{r-N}{r+N} \sigma_3.
\eee
Note that this is globally defined on TN since the radial function multiplying $\sigma_3$ vanishes at the nut. 
As discussed in \cite{JS1}, the coefficient $\tilde p$ is not quantised in the TN case, essentially because there is no non-contractible 2-sphere in TN.

\section{Gauged Laplace operator on TN, TB and ES geometries}

The Laplace operator associated to metrics of the  Bianchi IX form \eqref{genbianchi} is \cite{GM}
\bee
\Delta = \frac{1}{abcf}{\partial_r} \frac{abc}{f} \partial_r +  \frac{\xi_1^2}{a^2} +   \frac{\xi_2^2}{b^2} + \frac{\xi_3^2}{c^2},
\eee
where we write $\partial_r$ for $\partial/\partial r$. 
Gauging this operator  means  the minimal substitution of ordinary derivatives by covariant derivatives with respect to the connection $iA$ discussed above, i.e.
\bee
\partial_r \mapsto \partial_r + i A(\partial_r), \quad \xi_j\mapsto \xi_j+iA(\xi_j), \quad j=1,2,3.
\eee
Focussing on the TB family, the particular form 
 of the connection \eqref{genA}  and the duality relation $\sigma_i(\xi_j)=\delta_{ij}$ mean that  we only  need to substitute 
\bee
\label{TBsub}
\xi_3\mapsto \xi_3 + i \frac p 2\frac{(k+1)}{(k-1)}\frac{r-N}{r+N}.
\eee

The Laplace operator on the metric \eqref{genmet} for the TB family minimally coupled to the self-dual connection \eqref{genA} is therefore 
\begin{align}
\label{genlap} 
\Delta_k &= \frac{1}{r^2-N^2}\partial_r \left(r-kN\right)\left(r-\frac{N}{k}\right) \partial_r   + \frac{1}{r^2-N^2}(\xi_1^2 + \xi_2^2) \nonumber \\
&+ \frac{r^2-N^2}{4N^2\left(r-kN\right)\left( r-\frac{N}{k}\right)}\left(\xi_3+i\frac {p}{2}
\frac {k+1}{k-1}\frac{r-N}{r+N}\right)^2, \quad 1<k <\infty.
\end{align}
Treating the TN case analogously, and including the resulting gauged Laplace operator with connection \eqref{TNswitch}  in the  TB family  as the limiting case $k=1$, we define 
\begin{align}
\label{TNlap} 
\Delta_1 = \frac{1}{r^2-N^2}\partial_r \left(r-N\right)^2 \partial_r   + \frac{1}{r^2-N^2}(\xi_1^2 + \xi_2^2) + \frac{r+N}{4N^2(r-N)}\left(\xi_3+i\frac {\tilde p}{2}
\frac{r-N}{r+N}\right)^2\!\!\!\!.
\end{align}

To analyse the  spectrum of the family of Laplace operators \eqref{genlap} and of \eqref{TNlap} we exploit the symmetry of the TB  and TN spaces and  separate variables into a function  of the radial variable and  Wigner functions $D^j_{ms}$ on $SU(2)$.  With the conventions of \cite{VMK}, the latter have the form 
\bee
D^j_{ms}(\phi,\theta,\psi) =   e^{-im\phi}d_{ms}^j(\theta) e^{-is\psi},
\eee 
where  the functions $d_{ms}^j$ are often called  elements of the small $d$-matrix; their explicit form is given in \cite{VMK}. Here we only require that the functions $D^j_{ms}$ 
are eigenfunctions of the total angular moment, and the left and right actions of rotations about the 3-axis: 
\bee
(\xi_1^2 + \xi_2^2 + \xi_3^2) D^j_{ms}= -j(j+1)D^j_{ms}, \quad i\xi_3D^j_{ms}=sD^j_{ms},\quad i\partial_\phi
D^j_{ms}=mD^j_{ms}.
\eee 
The range of the labels is 
\bee
j\in \frac 12 \NN^0,  \qquad  s,m\in\{-j,-j+1,\ldots,j-1,j\},
\label{jsm}
\eee
and 
for later use we note that 
\bee
(\xi_1^2 + \xi_2^2) D^j_{ms}= (-j(j+1)+s^2)D^j_{ms}.
\eee 

We recall that in the TB case we need to check if the angular dependence is non-singular in the non-singular monopole gauge. If we use the Wigner function $D_{ms}^j$ in the singular gauge, this amounts to checking, according to \eqref{Phigauge}, if 
\bee
\label{localsect}
D_{msp}^{j\pm}(\phi,\theta,\psi) := e^{i\frac{p}{2}(\psi\mp\phi)} D^j_{ms} (\phi,\theta,\psi)= 
e^{-i\left(m\pm\frac p 2 \right)\phi}d^j_{ms}(\theta) e^{-i\left(s-\frac p 2 \right)\psi}
\eee
is smooth on the bolt (where $\psi$ is not defined) for $\theta >0$ ($+$) and $\theta <\pi$ ($-$). However, this happens precisely if $s=p/2$, in which case \eqref{localsect} is independent of $\psi$ and 
\bee
D_{m \frac p 2 p}^{j\pm}(\phi,\theta,\psi) = D^j_{m \frac p 2 }( \phi,\theta,\pm\phi)
\label{sectionreg}
\eee
which are a standard expressions for local sections, valid in the open sets   $\theta >0$ ($+$) and $\theta <\pi$ ($-$), of the line bundle of Chern class $p$ over the 2-sphere.  We conclude that  we can use Wigner functions when separating variables on the TB  and TN spaces equipped with the connections \eqref{genA} and \eqref{TNswitch}, but that in the TB case  we need to impose $s=p/2$ for any function which is non-zero at the bolt.  As we shall see, this is implemented rather naturally in our spectral analysis.

Inserting  the ansatz
\bee
\Phi(r,\phi,\theta,\psi) = u_{jms}(r) D^j_{ms}(\phi,\theta,\psi),
\eee
(no sum) into  the equation 
\bee
-\Delta_k\Phi = \epsilon\Phi
\eee
leads to  the radial  problem
\bee
\tilde H_k u_{jsm} = \epsilon u_{jsm},
\eee
with radial Hamiltonians, labelled by $k>1$,  defined  initially on $C^\infty((kN,\infty))$ and given by
\begin{align}
\label{genradlap}
\tilde H_k &= -\frac{1}{r^2-N^2}\frac{d}{dr} \left(r-kN\right)\left(r-\frac{N}{k}\right) \frac{d}{dr} + \frac{j(j+1)-s^2}{r^2-N^2} \nonumber \\
&\qquad \qquad\qquad \qquad \qquad \qquad+  \frac{r^2-N^2}{4N^2(r-kN)(r-\frac{N}{k})}\left(s-\frac p2\frac {k+1}{k-1} \frac{r-N}{r+N}\right)^2.
\end{align}
The condition $s=p/2$ which characterises a special situation in the angular dependence also leads to a special feature in the radial Hamiltonian. A cancellation in the last term in this case leads to the simplified form 
\begin{align}
\label{genradlapsp2}
\tilde H^{sp}_k &= -\frac{1}{r^2-N^2}\frac{d}{dr} \left(r-kN\right)\left(r-\frac{N}{k}\right) \frac{d}{dr} + \frac{j(j+1)-s^2}{r^2-N^2} \nonumber \\
&\qquad \qquad\qquad \qquad \qquad \qquad \qquad \qquad+\frac{s^2(r-kN) (r-N)}   {N^2(r-\frac{N}{k}) (k-1)^2 (r+N) }. 
\end{align}
The radial TN Hamiltonian, defined  initially on $C^\infty((N,\infty))$, is 
\bee
\tilde H_1 = -  \frac{1}{r^2-N^2}\frac{d}{dr} (r-N)^2 \frac{d}{dr}  + \frac{j(j+1)-s^2}{r^2-N^2} + \frac{r+N}{4N^2(r-N)}\left(s-\frac {\tilde p}{2}
\frac{r-N}{r+N}\right)^2 .
\eee
For completeness, we also include the radial Hamiltonian which arises in the study of the ES space. Using the conventions of  \cite{JS3}, it has the form 
\begin{align}
\label{ESham}
\tilde H_\infty& = -\frac{1}{r^2}  \frac{d}{d r}\left(r^2  - Lr \right)\frac{d }{d r} 
 + \frac{1}{r^2}\left(j(j+1) - \frac{p^2}{4}\right)  +
\frac{r}{r-L } \left(\frac{n}{2L}+\frac{p}{2}\left(\frac 1 L -\frac 1 r\right)\right)^2, \quad r\geq L.
 \end{align}
Here $L>0$ is a real parameter, $p,n\in \ZZ$ and $j = \frac{|p|}{2}, \frac{|p|}{2}+ 1, \frac{|p|}{2}+2,\ldots$.
We have denoted the Hamiltonian $\tilde H_\infty$ since it  can formally be obtained from $\tilde H_k$ of \eqref{genradlap} by identifying $L$  with the product $kN$ in the ES limit $k\rightarrow \infty$  \eqref{geolim} 
 together with the identification
  \bee
 \label{quantlim} 
 \left(s  -\frac {p}{2}\right) =  -\frac{n}{k}
 \eee
 for fixed $n\in\ZZ$.
For ease of reference, we summarise the range of the different parameters in the different models we consider in Table~\ref{params}.

\begin{table}
\begin{center}
\begin{tabular}{|c|c|c|c|c|c|c|}
\hline
Model & $k$  & N&  $p$  & $j$ & $s$ & $n$  \\
\hline TN & 1 &  $(0,\infty)$& $\tilde{p}\in\mathbb{R}$ &  $\frac12 \mathbb{N}_0$ &  $\{-j,-j+1,\ldots,j-1,j\}$ & ---  \\
TB & $(1,\infty)$ & $(0,\infty)$&  $\mathbb{Z}$ & $\frac12 \mathbb{N}_0$ &$\{-j,-j+1,\ldots,j-1,j\}$ &  ---  \\
ES & $\infty$ &0 & $\mathbb{Z}$ & $\frac{|p|}{2}+  \mathbb{N}_0$ & --- &  $\mathbb{Z}$ \\
\hline
\end{tabular}
\caption{Parameters of  the models considered in this paper. \label{params}} \end{center}
\end{table}


\section{Spectral properties of the radial Hamiltonian}
\label{rigorous}
\subsection{Standard form} 
\label{switch} In order to apply the results and techniques of \cite{BSSW} to the study of the  radial Hamiltonian
$\tilde H_k$ with $1\leq k\leq \infty$, we write it in standard Schr{\"o}dinger form via a unitary transformation. 
This requires, as a first step,  expressing it in terms of the proper radial distance from the bolt or nut. 
 Using the notation introduced in our discussion of the metric in \eqref{genmet}, we 
define the  proper radial distance 
\bee
\label{proper}
R(r)=\int_{kN}^r f(x) dx, \quad f(x)=\sqrt{\frac{x^2-N^2}{\left(x -kN\right) \left(x- \frac N k\right)}}.
\eee
We then rescale by   the function
\bee
\nu(r)= \sqrt{a^2c} = \sqrt{2N} \left(\left(r-kN\right)\left(r-\frac Nk\right)\left(r^2-N^2\right)\right)^{\frac 1 4}.
\eee
Noting that the radial derivative term in the Hamiltonian is 
\begin{align}
-\frac{1}{r^2-N^2}\frac{d}{dr} \left(r-kN\right)\left(r-\frac{N}{k}\right) \frac{d}{dr} 
= -\frac{1}{\nu^2} \frac{d}{dR} \nu^2 \frac{d}{dR} 
\end{align}
and that 
\bee
\label{flattening}
-\frac{1} {\nu^2} \frac{d}{dR} \nu^2 \frac{d}{dR} \left(\frac{\eta}{\nu}\right)=
 \frac{1}{\nu} \left( - \frac{d^2\eta }{dR^2} + \frac{1}{\nu} \frac{d^2\nu}{dR^2} \eta \right),
\eee
 the operators $\tilde H_k$ defined on $[kN,\infty)$ with measure $\nu^2  f dr$  are  unitarily equivalent to Hamiltonians of the standard form
\bee \label{widehatHk}
\widehat H_k = -\frac{d^2}{dR^2} + V^{\text{\tiny eff}}_{jsp},
\eee
defined on the half-line $[0,\infty)$ with measure $dR$ and effective potential
\bee
\label{Veff}
V^{\text{\tiny eff}}_{jsp}=\frac 1 \nu \frac{d^2 \nu}{d R^2}  +  \frac{j(j+1)-s^2}{r^2-N^2} + Q.
\eee
Here $Q$ combines the remaining non-derivative terms in  $\widehat H_k$ into a radial potential. It depends on the parameters $k,s$ and $p$ (respectively  $\tilde p$) summarised in Table~\ref{params}. Written in terms of $r$, we distinguish the following cases, making it explicit that there are important cancellations for some value of those parameters:
\bee
\label{Qcases}
Q(r)=\begin{cases} \frac{r+N}{4N^2(r-N)}\left(s-\frac {\tilde p}{2}
\frac{r-N}{r+N}\right)^2 \;  \; & \text{if} \; k=1,\  s\neq 0,\\
\frac{\tilde p^2}{16 N^2} \frac{r-N}{r+N} \;  \; & \text{if}  \; k=1,\  s=0, \\
\frac{ (r^2-N^2)}{4N^2(r-kN)(r-\frac{N}{k})}\left(s-\frac p 2\frac {k+1}{k-1} \frac{r-N}{r+N}\right)^2  \; & \text{if} \; k>1,\  s\neq \frac p2,  \\
\frac{p^2(r-kN) (r-N)}   {4N^2(r-\frac{N}{k}) (k-1)^2 (r+N) } \; & \text{if} \; k>1,\  s = \frac p 2 .
\end{cases}
\eee
We now describe the behaviour of the effective potential \eqref{Veff}  for small $r$ and large $R$.  In practice, we  used
\bee
\label{radprac}
\frac 1 \nu \frac{d^2 \nu}{d R^2}  = \frac {1} {f \nu} \frac{d }{d r}  \left(\frac 1 f
  \frac{d \nu}{ d r} \right).
\eee
to derive the asymptotic expansions in the following sections. 

\subsection{Short range asymptotics} 
 Defining 
 \bee
 \rho=r-kN,
 \eee
 we find the following expansions for small $\rho$ of the right hand side of \eqref{radprac},
 \bee
 \frac {1} {f \nu} \frac{d }{d r}  \left(\frac 1 f
  \frac{d \nu}{ d r} \right) = \begin{cases}\frac{3}{32N\rho}-\frac{1}{64N^2}  + \mathrm{o}  (\rho) &\text{if} \quad k=1, \\
   -\frac{1}{16kN \rho} + \frac{11}{16(k^2-1)N^2} + \mathrm{o}  (\rho) &\text{if} \quad k>1.
\end{cases} 
    \eee
Now, still  for  small  $\rho$, the definition \eqref{proper} implies that 
 \bee
 R = \begin{cases} 
 2\sqrt{2N\rho} +\mathrm{o}(\rho^{\frac 32 }) &\text{if} \quad k=1, \\
  2\sqrt{kN\rho} +\mathrm{o}(\rho^{\frac 32}) &\text{if} \quad k>1,
 \end{cases}
 \eee
 or 
 \bee
 \rho= \begin{cases} 
 \frac{R^2}{8N}  +\mathrm{o}(R^4) &\text{if} \quad k=1, \\
  \frac{R^2}{4kN}  +\mathrm{o}(R^4) &\text{if} \quad k>1.
 \end{cases}
 \eee
 These expressions also hold for the ES limit \eqref{geolim}. 
Hence 
 \bee
 \label{34result}
 \frac 1 \nu \frac{d^2 \nu}{d R^2} = \begin{cases}\frac{3}{4R^2}-\frac{1}{64N^2}  + \mathrm{o}  (R^2) &\text{if} \quad k=1, \\
   -\frac{1}{4R^2} + \frac{11}{16(k^2-1)N^2} + \mathrm{o}  (R^2) &\text{if} \quad k>1,
\end{cases} 
    \eee
  We also have the expansion of the $j$-dependent term
\bee
 \frac{j(j+1)-s^2}{r^2-N^2}=  \begin{cases}
  \frac{j(j+1)-s^2}{2N\rho} +  \mathrm{o}  (1) &\text{if} \quad k=1, \\
 \frac{j(j+1)-s^2}{(k^2-1)N^2}   +  \mathrm{o}  (\rho)  &\text{if} \quad k>1,
   \end{cases}
\eee
which turns into
\bee
\label{jjresult}
 \frac{j(j+1)-s^2}{r^2-N^2}= \begin{cases}
  4\frac{j(j+1)-s^2}{R^2} +  \mathrm{o}  (1) &\text{if} \quad k=1, \\
   \frac{j(j+1)-s^2}{(k^2-1)N^2}   +  \mathrm{o}  (R^2)  &\text{if} \quad k>1.
   \end{cases}
   \eee
To obtain the corresponding  results for the ES case we need to combine the limits
\eqref{geolim} and \eqref{quantlim}, so replace $(k^2-1)N^2\mapsto L^2$ and $s\mapsto \frac p2$ in the above expressions.

Finally turning to  $Q$,  for $k>1$ we obtain
 \bee
 \label{Qcent}
 \frac{ (r^2-N^2)}{4N^2(r-kN)(r-\frac{N}{k})}\left(s-\frac p 2\frac {k+1}{k-1} \frac{r-N}{r+N}\right)^2
 = \frac{k^2}{R^2} \left(s-\frac p 2\right)^2 + \mathrm{o}(1),
 \eee
 with the ES limit given by the replacement $k^2(s- p /2)^2 \mapsto n^2$. 
 For $k=1$ we find
  \bee
  \frac{r+N}{4N^2(r-N)}  \left(s-\frac {\tilde p}{2} \frac{r-N}{r+N}\right)^2
  =  \frac{4s^2}{R^2} +\mathrm{o}(1).
  \eee
We can combine  this term with the terms \eqref{34result}  and  \eqref{jjresult} to write  the centrifugal potential in the $k=1$  case as 
 \bee
 \frac{\frac 34 + 4j(j+1)}{R^2} = \frac{(2j+1)^2 -\frac 14 }{R^2}.
 \eee
 Our results are summarised  in the second column of Table~\ref{asypots}.

\subsection{Long range asymptotics}
The computation of the long range asymptotic behaviour is a little messy, but straightforward in principle. Looking at the form of  $Q$  when $k>1$  and $s\neq \frac p2$ in \eqref{Qcases}, for example, one finds 
\begin{align}
 &\frac{ (r^2-N^2)}{4N^2(r-kN)(r-\frac{N}{k})}\left(s-\frac p 2\frac {k+1}{k-1} \frac{r-N}{r+N}\right)^2 =
 \frac {1} {4N^2} \left(s-\frac p2\frac {k+1}{k-1}\right)^2 \qquad \qquad\qquad \qquad \qquad \nonumber \\
& \qquad \qquad + \frac{(k+\frac 1 k)}{4N}\left( s-\frac {p}{2}\frac {k+1}{k-1} \right) \left(s-\frac {k+1}{k-1}\left( 1-
\frac{4k}{k^2+  1} \right)\frac {p}{2}\right)
\frac 1 r + \mathrm{o}\left(\frac {1}{r^2} \right).
\end{align}
Using  the integral defining $R$ to determine the asymptotic relation between $R$ and $r$,
\bee
R=  r+  \frac 12  \left( k + \frac 1 k \right) N \ln r  + \mathrm{o}(1),
\eee
we deduce 
\begin{align}
 &\frac{ (r^2-N^2)}{4N^2(r-kN)(r-\frac{N}{k})}\left(s-\frac p 2\frac {k+1}{k-1} \frac{r-N}{r+N}\right)^2 =
  \frac {1} {4N^2} \left(s-\frac p2\frac {k+1}{k-1}\right)^2 \qquad \qquad\qquad \qquad \qquad \nonumber \\
& \qquad \qquad + \frac{(k+\frac 1 k)}{4N}\left( s-\frac {p}{2}\frac {k+1}{k-1} \right) \left(s-\frac {k+1}{k-1}\left(1-
\frac{4}{k^2+ 1 } \right) \frac {p}{2}\right)
\frac 1 R + \mathrm{o}\left(\frac {1}{R} \right). 
\end{align}
One checks that the case $s=\frac p2$ has the same large-$R$ asymptotics.
Expanding $Q$ similarly for $k=1$ and using
\bee
\nu^2= 2N\sqrt{(r-kN)(r-\frac N k) (r^2-N^2)} =2Nr^2\left(1-\frac 12 \left( k + \frac 1 k \right) \frac N r\right)+\mathrm{o}(1)
 \eee
to evaluate \eqref{radprac} finally leads to the asymptotic expansions for the effective potential   summarised in the third column of Table~\ref{asypots}.

\begin{table}[ht]
\centering
\begin{tabular}{|c|ll|rl|}
\hline
\text{Model} & \quad $V^{\text{\tiny eff}}_{jsp} $ for $R\to 0$ & & &$V^{\text{\tiny eff}}_{jsp}$ for $R\rightarrow \infty$ \\ \hline &&&& \\
TN & $[(2j+1)^2 -\frac 14 ]\frac{1}{R^2}$ & $+\mathrm{o}\left(1 \right)  $ &  $\frac{\left(s-\frac {\tilde p}{2}\right)^2}{4N^2} +$ & $\frac{\left( s-\frac {\tilde p}{2} \right)\left( s+ \frac {\tilde p}{2} \right)}{2N}\,\, \frac 1 R$ \\ $(k=1)$ &&&$+\mathrm{o}\left(\frac{1}{R} \right) $ & \\ &&&& \\
 TB & $\left[ k^2(s\!-\!\frac p2 )^2-\frac 14\right]\!\frac{1}{R^{2}}$ &
$+\mathrm{o}(1) $ & $\frac {\left(s-\frac p2\frac {k+1}{k-1}\right)^2}{4N^2}+$ & $\frac{(k+ \frac 1 k)\!\left(\! s-\frac {p}{2}\frac {k+1}{k-1}\! \right) \!\left[\!s-\frac {p}{2} \frac {k+1}{k-1}\!\left(\!1-
\frac{4k}{k^2+  1} \!\right) \!\right]}{4N}\,\,\frac 1 R$ \\ $(1<k<\infty)$ &&&$+\mathrm{o}\left(\frac{1}{R} \right) $ & \\ &&&& \\
ES &$(n^2-\frac 1 4)\frac{1}{R^2}$ & $+\mathrm{o}(1)$ & 
$\frac{(n+p)^2}{4L^2}+$ &$\frac{1}{4L(n+p)(n-p)}\,\,\frac 1 R$ \\ $(k=\infty)$ &&&$ +\mathrm{o}\left(\frac{1}{R} \right) $ &\\
&&&&\\
\hline
 \end{tabular}
\caption{Asymptotic behaviour of the effective potentials. The ranges of the various parameters are given in  Table~\ref{params}.}
\label{asypots}
\end{table}

\begin{figure}[!ht]
  \includegraphics[width=\linewidth]{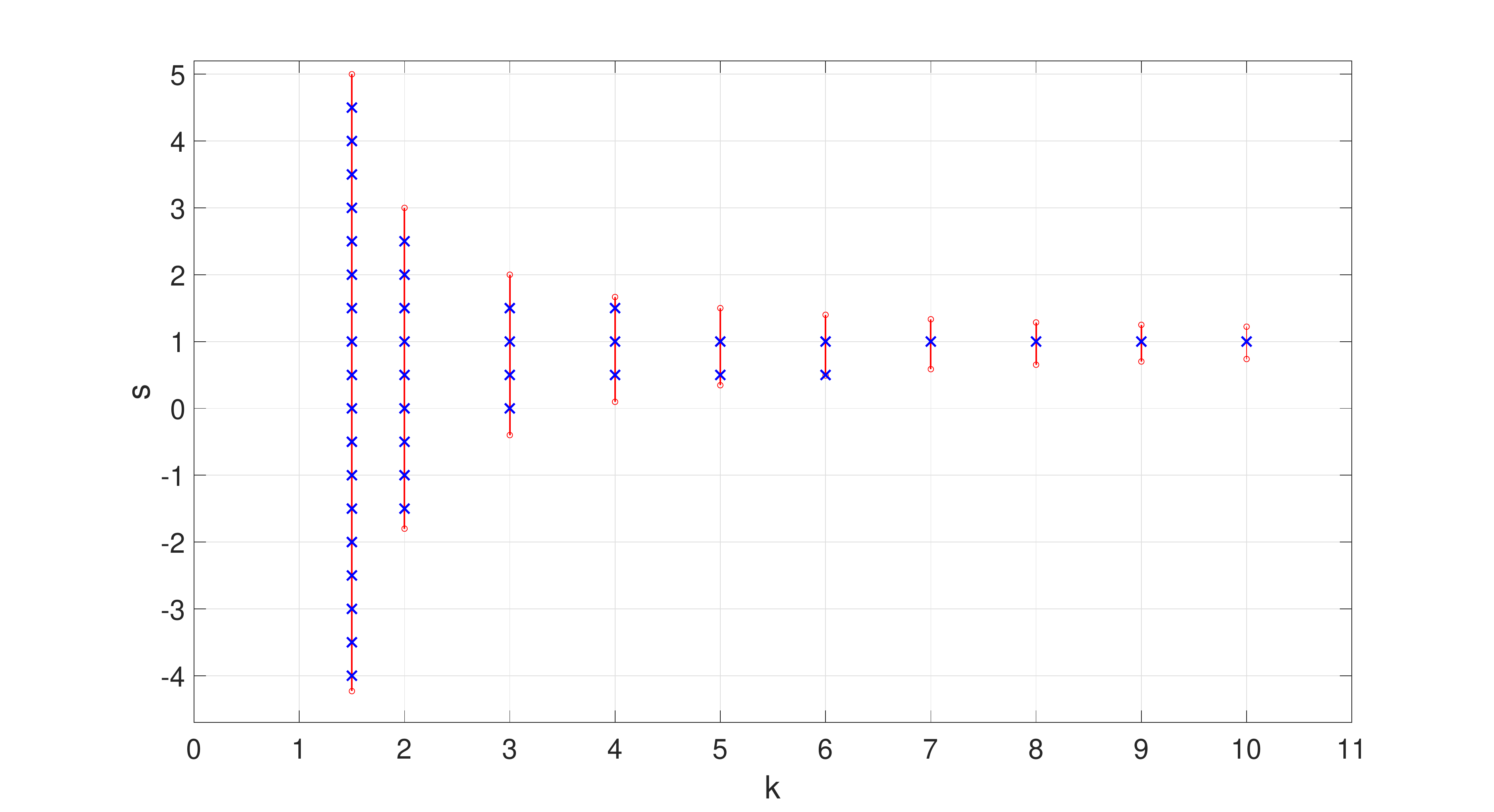}
  \caption{Values of  $s$  that allow  for bound states according to \eqref{boundcond} when $p=2$.}
  \label{fig:sintp2}
\end{figure}

The results collected in Table~\ref{asypots} show that the effective potential typically has a $1/R^2$ singularity at $R=0$ with a coefficient $\geq -\frac 14$ while the  leading large $R$ terms in Table~\ref{asypots}   are attractive  Coulomb ($1/R$)  potentials  provided
the following conditions are satisfied:
\begin{align} 
\label{boundcond}
k=1 &\qquad:\qquad  
-\frac{|\tilde p|}{2} < s < \frac {|\tilde p|}{2},  \nonumber  \\
1<k < \infty ,\, p\geq 0 &\qquad:\qquad   
\frac {k+1}{k-1}\left(1-
\frac{4k}{k^2+   1 } \right) \frac {p}{2} < s <\frac {p}{2} \frac {k+1}{k-1},  \nonumber  \\
1<k < \infty , \,p<0 & \qquad:\qquad  
\frac {p}{2} \frac {k+1}{k-1} < s <  \frac {k+1}{k-1}\left(1-
\frac{4k}{k^2+   1 } \right) \frac {p}{2},  \nonumber  \\
k=\infty &\qquad:\qquad   -|p| < n < |p|.
\end{align}
When $k>1$,   we have $\frac {k+1}{k-1}\left(1- \frac{4k}{k^2+   1 }\right)<1$, so that the allowed interval for an attractive Coulomb potential always includes $s=\frac p2$. In particular, it is never empty. We illustrate the  values of $s$  that lead to Coulomb attraction  for given $p$ as a function of $k$ in Figure~\ref{fig:sintp2}.  

The conditions for attractive Coulomb tails in the TN and ES cases were previously derived in \cite{JS2} and \cite{JS3} and also interpreted geometrically. As pointed out there, the existence of bound orbits and bound states can be understood qualitatively in terms of  the motion of a charged particle on certain two-dimensional submanifolds of the four-geometries. In terms of the coordinates introduced after \eqref{genbianchi}, these submanifolds
are the  disks with angular coordinate $\psi$ and radial coordinate  $c$ at fixed values of $(\theta,\phi)$. The dynamics on these disks is that of  a charged particle moving on a disk under the influence of a transverse magnetic field (stemming from the $U(1)$ connection), much studied in the context of Landau levels. The particle moves in circular orbits whose radius is proportional to the angular momentum, but in order to fit onto the disk the angular momentum has to satisfy a  bound set by the strength of the magnetic field. For the motion on TN and ES, the magnetic field strength  is proportional to $\frac{\tilde p}{2}$ and $p$, and the relevant angular momenta are  $s$ and $n$, resulting in the  conditions  \eqref{boundcond}. For details and further references  see \cite{JS2}, in particular the discussion of a toy model in Section 2 of that paper. The  condition for bound states in the TB family depends on the  geometrical parameter $k$ and is more intricate, but the qualitative picture is as for TN and ES. 
 
\subsection{Domain, selfadjointness and spectrum}

We now state and prove the central results of this paper. Their validity is a consequence of the fact that, irrespective of the model, the effective potentials \eqref{Veff} in the radial reduction of the Hamiltonian turns into a one-dimensional Schr{\"o}dinger operator which has a Calogero-type singularity at $R=0$ and a Coulomb-type behaviour at $R=\infty$.

We begin by describing the domain of a selfadjoint realisation of the radial Hamiltonian \eqref{widehatHk}. Let 
\[
     m_k=\begin{cases} 2j+1 & k=1 \\
              k(s-\frac{p}{2}) & k\in(1,\infty) \\
n & k=\infty \end{cases},
\]
so that the Calogero contribution of $V^{\mathrm{eff}}$ at $R=0$ has coefficient $m_k^2-\frac14$. Since $m_k^2\geq 0$, the differential expression $\widehat{H}_k$ is symmetric and semi-bounded below on the domain $C^\infty_0((0,\infty))$ of compactly supported functions, see \cite[Lemma~1]{BSSW}. In fact we know that it is always bounded below by zero, from the positivity of the Laplacian \eqref{genbianchi} associated to a Riemannian metric.  The operator  $\widehat{H}_k$ is limit point at  $R=\infty$ (so no boundary condition is required there) but the  value of $m_k$ determines the nature of  the operator at $R=0$ and hence its domain of selfadjointness. To see this, consider
\bee 
H=-\frac{d^2}{dR^2}+\frac{m_k^2-\frac14}{R^2}
\eee 
on its minimal domain $\mathcal{D}_{\mathrm{min}}\subset L^2(\mathbb{R}^+)$,
which is found from the closure of the operator defined on $C^\infty_0((0,\infty))$. Let $\chi\in C^\infty([0,\infty))$ be a mollifier identically equal to 1 in $[0,1]$ and vanishing in $[2,\infty)$ and set $\eta_{m_k}(R)=\chi(R)R^{\frac12+m_k}$. Defining
 \[
    \mathcal{D}=\mathcal{D}_{\mathrm{min}}+\eta_{m_k} \mathbb{C},
\]
we see that $\mathcal{D}=\mathcal{D}_{\mathrm{min}}$ for $m_k\geq 1$. According to \cite[Proposition~4.17]{BDG2011}, it is known that the operator
$H$ with domain $\mathcal{D}$ is selfadjoint. The following lemma shows that the operator $\widehat{H}_k$ extended to $\mathcal{D}$ also has this property.

\begin{lemma} \label{domain}
The operator $\widehat{H}_k:C^\infty_0((0,\infty))\longrightarrow L^2(\mathbb{R}^+)$ is  essentially selfadjoint, when
\begin{enumerate}
\item $k=1$ (TN space), or
\item  $1<k<\infty$ (TB family) and $m_k\geq 1$, or
\item $k=\infty$ (ES space) and $n\not=0$.
\end{enumerate}
Otherwise, it  has deficiency indices $(1,1)$. Moreover, the extension $H_k$ of $\widehat{H}_k$ to $\mathcal{D}$ is a selfadjoint operator.
\end{lemma}

\textbf{Proof.}
It is classically known that $H\upharpoonright_{C^{\infty}_0((0,\infty))}$ is essentially selfadjoint if and only if $m_k\geq 1$. Moreover, for $0\leq m_k<1$, it has deficiency indices $(1,1)$. According to Table~\ref{asypots} the condition for $m_k\geq 1$ holds always for TN and whenever $|n|\geq 1$ for ES.

Now, according to \cite[Lemma~4]{BSSW},  and  with  the following notation for the constant term in the long range asymptotic (c.f. Table~\ref{asypots}),
 \begin{align} \label{essspecC0}
C_0= \begin{cases} \frac{ 1} {4N^2} \left(s-\frac {\tilde p}{2}\right)^2  \;  & \text{for} \; k=1, \\
\frac {1} {4N^2} \left(s-\frac p2\frac {k+1}{k-1}\right)^2 \;   & \text{for}\;  1<k<\infty, \\
\frac{(n+p)^2}{4L^2} \; &  \text{for} \; k=\infty,
\end{cases}
\end{align}
the potential $V^{\mathrm{eff}}(R)-\frac{m_k^2-\frac14}{R^2}-C_0$  is relatively compact with respect to $H$. Hence it is relatively bounded with bound 0 with respect to $H \upharpoonright_{C^{\infty}_0((0,\infty))}$.
Thus the Kato-Rellich theorem ensures the conclusions stated in the lemma.\hfill $\Box$

\medskip

We now characterise the spectrum of $H_k$.
\begin{theorem} Let $C_0$ be defined as in \eqref{essspecC0}. Then 
\begin{equation}
   \operatorname{Spec}_{\mathrm{ess}}(H_k)=[C_0,\infty).
\end{equation}
Moreover, $H_k$ has infinitely many eigenvalues below $C_0$ provided the conditions \eqref{boundcond} hold.
\end{theorem}
\textbf{Proof.} The claim about the essential spectrum is a consequence of 
\cite[Lemma~4]{BSSW}. Moreover, under the hypotheses \eqref{boundcond}, the coefficient of the long range Coulombic part of the potential is negative and so the claim about the discrete spectrum follows from \cite[Theorem~6]{BSSW}.
This ensures the validity of the theorem. \hfill $\Box$

The condition  $s=p/2$ plays a special role in the  spectral analysis of the TB family and deserves a comment. In our discussion of  sections of the line bundle over the bolt  used in separating variables,   this emerged as a regularity condition for the angular dependence on the bolt, see the paragraph following equation \eqref{localsect}.  With  the selfadjoint extension $H_k$  we  now find that eigenfunctions of the gauged Laplace operator of the TB family are non-vanishing at the bolt precisely when the angular dependence is regular, i.e. when $s=p/2$.  To see this, recall  from Section~\ref{switch} that  $\eta/\nu $ is an eigenfunction of the radial Hamiltonian $\tilde H_k$ if $\eta $ is an eigenfunction of $\widehat{H}_k$. Near $R=0$, the leading term of the latter is  $\mathrm{O}(R^{\frac 12+m_k})$. With $\nu =\mathrm{O}(\sqrt R )$ near $R=0$, the eigenfunction $\eta/\nu$ has leading term $R^{m_k}$, which, in the TB case,  is non-vanishing at the bolt  precisely if $s=p/2$.  This is a more involved version of what happens when studying the Laplace operator in the plane in terms of polar coordinates $(\rho,\varphi)$, see also the discussion of Lemma 4  in \cite{BSSW}.  The angular functions $e^{in\varphi}$ for integer $n$ are ill-defined at the origin except when $n=0$. With the right choice of selfadjoint extension of the radial Laplace operator, briefly reviewed above, eigenfunctions  are non-vanishing (but finite) at the origin if and only if $n=0$, thus ensuring that eigenfunctions  obtained via separating variables are in the domain of the planar Laplacian.  

It is natural to wonder what happens when the coefficient of the Coulomb contribution in the large $R$ regime  is non-negative, i.e. in the complement of the intervals given by  \eqref{boundcond}. When the Coulomb term is repulsive, we necessarily have $C_0>0$ and we cannot rule out a discrete spectrum below $C_0$. However, the neutral case (vanishing Coulomb term) is more subtle. 
 For $k=1$, $s=\frac{\tilde{p}}{2}$ implies $C_0=0$ while $s=-\frac{\tilde{p}}{2}$ implies $C_0>0$. For $1<k<\infty$, $s=\frac{p}{2} \frac{k+1}{k-1}$ implies $C_0=0$ while $s=\frac{p}{2} \frac{k+1}{k-1}\left(1-\frac{4k}{k^2+1}\right)$ implies $C_0>0$. Finally, for $k=\infty$, $n=-p$ implies $C_0=0$ while $n=p$ implies $C_0>0$. For $C_0=0$, we know from $\operatorname{Spec}(H_k)=[0,\infty)$ that there is no discrete spectrum in this case. However,  when the Coulomb  contribution vanishes while $C_0 >0$  we are once again not able to draw any definite conclusion.

\section{Approximating the eigenvalues}

\subsection{Exact and numerical eigenvalues}

Having established the existence of infinitely many eigenvalues for the gauged Laplacians of the TN, TB and ES geometries, we now turn to their numerical computation. The TN metric is special compared with the other two geometries  in that both the associated classical problem (geodesics) and the associated quantum problem (spectrum of the Laplacian) can be solved analytically,  with simple algebraic expressions for the eigenvalues. This remains true in the presence of the magnetic field \eqref{TNswitch} and is explored in detail in  \cite{JS2}. There is no known method for obtaining exact eigenvalues of  the TB or ES Laplacians, so  numerical schemes are required  to investigate these two cases. The ES case was examined in \cite{JS3}, via an implementation of the shooting method. The main purpose of this section is  the approximation of the eigenvalues for the TB family by means of a projection method, but we begin by presenting  an analytical approximation of  the eigenvalues in terms of TN eigenvalues.

\subsection{The Taub-NUT approximation}

The starting  point of this approximation  is the asymptotic form of the TB Laplacian. This turns out to be of the TN form and its spectral properties can therefore be determined exactly. In this way we  obtain an analytic approximation to the TB spectrum which we call the TN approximation. Although we have not been able to justify the validity of this approximation rigorously, we show evidence that it is  in close agreement with our numerical results.

 Following \cite{JS2}, the radial eigenvalue equation  for the TN Laplacian  coupled to \eqref{TNswitch}, expressed in the isotropic coordinates \eqref{TNisotropic} is  
 \bee
 \label{TNreference}
- \frac{d^2\eta}{d \rho^2} +\left[
 \frac{ \tilde j (\tilde j +1)}{\rho^2} + \frac{2 \tilde s \left(\tilde s -\frac{\tilde p}{2}\right)}{\Lambda \rho} +\left(\frac{\tilde s -\frac{\tilde p}{2}}{\Lambda} \right)^2\right]u= E \left(1+\frac \Lambda \rho\right) \eta, \quad \rho \geq 0.
\eee
We have added tildes to the angular momentum  eigenvalues $j$ and $s$  to avoid a clash in the notation.  The eigenvalues $E$ exist when the condition \eqref{boundcond} for $k=1$ holds. The entire spectrum is given by \cite{JS2}
\bee
\label{Espectrumm}
E(\tilde p,\tilde s,\tilde n) = \frac{2}{\Lambda^2} \left(  \tilde s\left(\tilde s - \frac{ \tilde p }{2}\right)  +  \tilde n\sqrt{\tilde n^2 -\left(\tilde  s - \frac{\tilde p}{2}\right)\left(\tilde  s + \frac{\tilde p}{2}\right)} - \tilde n^2\right),
\; \tilde n  =  |\tilde s|+1, |\tilde s|+2,  \ldots.
\eee
An important aspect of this formula is that $j$  only restricts the range of $s$ according to \eqref{jsm} but does not enter explicitly.
This is consequence of a dynamical symmetry of the spectral problem which manifests itself through  additional operators  of Runge-Lenz type which commute with the Hamiltonian \cite{JS2}.  It matters in the current context because it implies that the spectrum is determined by the coefficients of the terms in the eigenvalue equation \eqref{TNreference} which are  $\mathrm{O}(1), \mathrm{O}(1/\rho),\mathrm{O}(E)$ and $\mathrm{O}(E/\rho)$. 

We now expand the radial eigenvalue problem for the TB Hamiltonian  $\tilde H_k$  \eqref{genradlap}  to extract the terms of this order. 
For this purpose,  we introduce abbreviations 
\bee
F(r) = \frac{r^2-N^2}{(r-kN)\left(r-\frac{N}{k}\right)},  \quad G(r)=
s-\frac p2\frac {k+1}{k-1} \frac{r-N}{r+N},
\eee
for the functions which appear in \eqref{genradlap}, and multiply the radial eigenvalue problem  by $F$. The resulting  equation   
\bee
F  \tilde  H_k u=EFu
\eee
is equivalent to
\bee
\label{firstversion}
-\frac{1}{\mu^2} \frac{d}{dr} \left( \mu^2 \frac{du}{dr}\right)   + \frac{j(j+1)-s^2}{r^2-N^2}u +\frac{F^2G^2}{4N^2}u=EFu,
\eee
where $\mu^2= (r-kN)\left(r-\frac{N}{k}\right)$. We can use the identity \eqref{flattening} to deduce that  \eqref{firstversion} can be written as 
\bee
\label{secondversion}
- \frac{d^2\eta}{dr^2}   + \frac{1}{\mu}\frac{d^2\mu}{dr^2} \eta+ \frac{j(j+1)-s^2}{r^2-N^2} \eta+\frac{F^2G^2}{4N^2}\eta=EF\eta,
\eee
for $\eta= u/\mu$. In this formulation, the TB eigenvalue equation can be compared with the TN eigenvalue equation \eqref{TNreference}.  To match the range of the radial variable in the TN problem, we set $\rho = r-kN$,  and think of  $F$ an $G$ as functions of $\rho$. Then we expand 
\bee
F(\rho) = \frac{(\rho +(k+1) N)(\rho +(k-1)N)}{\rho \left(\rho + \left(k-\frac 1 k\right)N\right)} = 1 + \frac{N\left(k+\frac  1 k\right)}{\rho} +\mathrm{O}\left( \frac{1}{\rho^2}\right),
\eee
and 
\bee
G(\rho) =  \left(s-\frac p 2\frac {k+1}{k-1} \frac{\rho +(k-1)N}{\rho +(k+1)N}\right) = \left( s-\frac p 2 \frac{k+1}{k-1}\right) + p\frac{k+1}{k-1} \frac N\rho  + \mathrm{O} \left( \frac{1}{\rho^2}\right),
\eee
to deduce 
\begin{align}
&  \frac{1}{4N^2}  F^2(\rho) G^2(\rho) = \frac{1}{4N^2}   \left( s-\frac p 2 \frac{k+1}{k-1}\right)^2      \qquad  \nonumber \\ & \qquad \qquad
+  \frac{1}{2N}\left( s-\frac p 2 \frac{k+1}{k-1}\right) \left(\left(k+ \frac 1 k\right) s + \frac p 2\frac{k+1}{k-1} \left(2- k-\frac 1 k\right)\right) \frac{1}{\rho} 
 + \mathrm{O}\left( \frac{1}{\rho^2}\right).
\end{align}
The remaining terms in \eqref{secondversion} are $\mathrm{O}\left(  1/ \rho^2 \right)$. Although we do not need the coefficient of the $1/\rho^2$ term for the computation of the TN approximation, we include it for completeness. Setting 
\bee
J^2 =  j(j+1)+ \frac{(k + 1)(k - 1)\left(s -\frac p 2  \frac{k + 1}{k-1}\right)\left( (k^2-3)s - \left(k^2 -2 k +  3  \right)\frac p 2 \right)}{4k^2},
\eee
 the TN approximation  to  the TB radial equation is 
\begin{align}
\label{thirdversion}
- \frac{d^2\eta}{d\rho^2}    + \frac{J^2}{\rho^2} \eta + \frac{1}{2N}\left( s-\frac p 2 \frac{k+1}{k-1}\right) \left(\left(k+ \frac 1 k\right) s + \frac p 2\frac{k+1}{k-1} \left(2- k-\frac 1 k\right)\right) \frac{1}{\rho}  \eta \nonumber \\
+ \frac{1}{4N^2}   \left( s-\frac p 2 \frac{k+1}{k-1}\right)^2\eta=
E \left(1 + \frac{N\left(k+\frac  1 k\right)}{\rho}\right)\eta.
\end{align}
Comparing with \eqref{TNreference} we deduce that by setting 
\bee
\Lambda = \left(k+\frac  1 k\right)N,
\eee
as well as 
\bee
\label{sptrans}
 \tilde p = p\frac{k+1}{k-1},  \quad \tilde s = \frac{\tilde p}{2}  +\frac 1 2 \left(k+ \frac 1 k\right)  \left( s-\frac p 2 \frac{k+1}{k-1}\right), 
\eee
we obtain the  TN approximation to the   energy levels  from \eqref{Espectrumm}. It is convenient to set  $\sigma = \tilde s -\frac{\tilde p}{2}$ and to compute  the TN approximation from
\bee
\label{TNspectrum}
\varepsilon(p,s,\tilde n) = \frac{ 1 }{N^2\left(k+ \frac 1 k\right)^2 }
\left( 2\sigma^2  + \tilde p \sigma +  2 \tilde  n\sqrt{\tilde n^2 -\left( \sigma^2 + \tilde p \sigma \right)} - 2\tilde n^2\right),
\; \tilde  n  =  |\tilde s|+1, |\tilde s|+2,  \ldots .
\eee

Substituting the parameters $N, k, s,p$ of a given TB  radial Hamiltonian into \eqref{sptrans} and \eqref{TNspectrum}  gives the TN approximation to the eigenvalues of that Hamiltonian which we compare with numerically computed eigenvalues below.
As  a consistency check we compared the conditions for bound states in the TN problem derived in \cite{JS2}, and found that it  agrees with the condition derived from \eqref{boundcond} for the case $1<k<\infty$.

It is instructive to determine the dependence of the eigenvalues on $\tilde n$ as $\tilde{n}\to \infty$, and to compare with the standard Coulomb problem. Firstly, note that 
\[
\sigma +\tilde p =\frac 12 \left(k+\frac 1k\right) \left(s-\frac {k+1}{k-1}\left(1-
\frac{4k}{k^2+   1 } \right) \frac {p}{2}\right).
\]
Hence,
\begin{align}
\label{Espectrumgenapprox}
\varepsilon(p,s,\tilde n) & = \frac{1}{4N^2}\left(s-\frac p2 \frac{k+1}{k-1}\right)^2 - \frac{(\sigma(\sigma +\tilde p))^2}{4N^2\left(k+\frac 1 k\right)^2} \frac{1}{\tilde n^2} + \mathop{O}\left(\frac{1}{\tilde n^4}\right)
\nonumber \\
&=\frac{1}{4N^2}\left(s-\frac p2 \frac{k+1}{k-1}\right)^2 \nonumber \\
& -\frac{\left(k+\frac 1 k\right)^2}{64N^2}\left( s-\frac {p}{2}\frac {k+1}{k-1} \right)^2 \left(s-\frac {k+1}{k-1}\left(1-\frac{4k}{k^2+   1 } \right) \frac {p}{2}\right)^2 \frac{1}{\tilde n^2} + \mathop{O}\left(\frac{1}{\tilde n^4}\right).
\end{align}
Thus,  the leading terms agrees with the lower bound on the essential spectrum for the TB family given in \eqref{essspecC0}, 
and the coefficient of the eigenvalue  spacing term $1/\tilde n^2$ is precisely a quarter of the  square of  the coefficient of the Coulomb potential in Table~\ref{asypots}, which is a familiar feature  of the standard Coulomb problem in three dimensions.

\subsection{Numerical results}  
\label{numerical}

For the computation of eigenvalues of the radial Hamiltonians \eqref{genradlap}(when $s\neq p/2$)  and \eqref{genradlapsp2} (when $s=p/2$)  in the TB family, we have set the parameter $N$ in \eqref{Upot} to $N=1$, but consider a range of benchmark values for $k$. We numerically compute upper bounds on the eigenvalues using  the Riesz-Galerkin method  via the Matlab toolbox Chebfun which implements a Chebyshev basis, see \cite{TBD} and references therein. This requires a truncation of the interval $[r_+,\infty)$ to a finite interval which avoids  singularities at $r_+$. In practice we work on $[r_+ + h, m]$ for suitable small $h$ and suitably large $m$. For a detailed discussion of the method in a similar context see \cite{BSSW}.

\begin{figure}[!ht]
  \includegraphics[width=\linewidth]{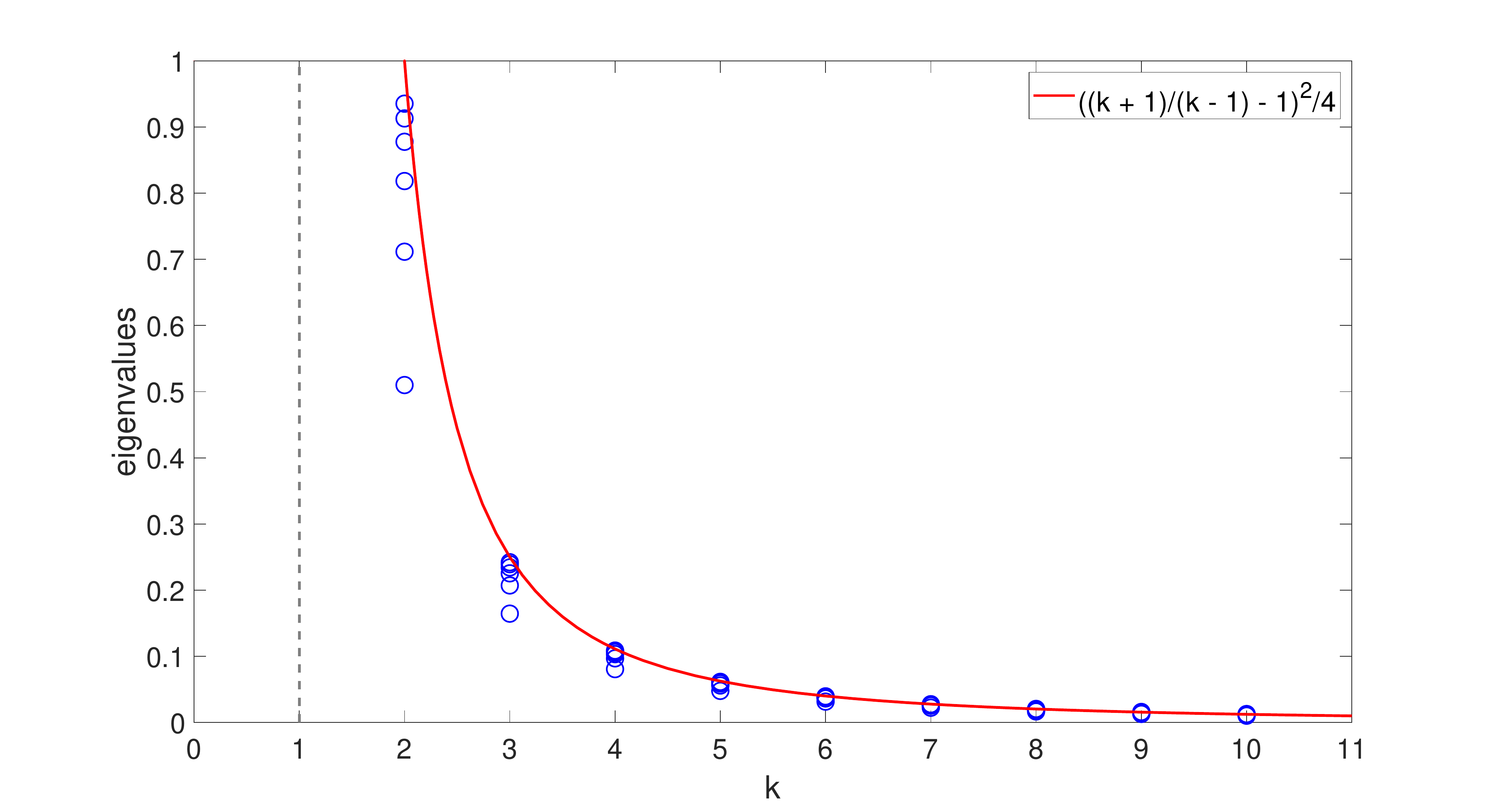}
  \caption{The numerically computed eigenvalues as a function of $k$  when $(s,j,p)=(1,1,2)$, with the bottom of the essential spectrum $C_{0}$ shown as a continuous curve.}
  \label{fig:plot12}
\end{figure}

We illustrate the  dependence of the  numerically computed  eigenvalues on the geometrical parameter $k$ 
in Figure~\ref{fig:plot12}, where we plot their numerical approximation together with the lower end of the essential spectrum $C_{0}$ for the case $(s,j,p)=(1,1,2)$ (still keeping $N=1$).

In Table~\ref{tab:data12} we show the numerically computed six lowest eigenvalues $\lambda_1,\ldots,\lambda_6$  for the four cases $(s,j,p)=(1,1,2),(1,2,2),(0,2,2),(0,0,2)$ respectively. We compare the answer with the TN approximation formula, and compute the relative error  according to 
\bee
RE_{l}=\frac{\vert \lambda_{l}-\lambda^{TN}_{l}\vert}{C_{0}-\lambda_{l}},
\label{relerror}
\eee
where $\lambda_{l}$ is the numerically calculated  $l$-th eigenvalue, $\lambda^{TN}_{l}$ is the TN approximation for the $l$-th eigenvalue computed according to \eqref{TNspectrum}, 
and $C_{0}$ is the lower bound of  the essential spectrum given in  \eqref{essspecC0}. Some of our numerically computed eigenvalues lie above $C_0$. By construction, the Riesz-Galerkin method does  not give any information about  eigenvalues embedded in the essential spectrum \cite[Section~4.5]{D}. Where our 
 numerical results lie above the threshold of the essential spectrum we have
therefore  discounted them.

The data collected in Table~\ref{tab:data12} shows several  trends regarding the agreement between  the TN approximation and  the numerically computed eigenvalues.
\begin{itemize}
\item As already seen in \cite{JS3} for ES,  the   agreement is best  when $j=s$. In two such cases, namely $k=2$ with $(j,s,p)=(1,1,2)$ and $k=3$ with  $(j,s,p)=(0,0,2)$ it is consistently very useful, with an accuracy better than $8\%$. When $|s| < j$, the TN approximation is not useful.
\item The agreement improves with the order of the eigenvalue. This is expected since eigenfunctions for higher eigenvalues have more of  their support  in the asymptotic region of TB where, by construction,  the  Hamiltonian is close to its TN approximation. 
There is an increase in the relative error in  the cases directly before where our numerical approximation become unreliable and lies above the threshold of the essential spectrum. This indicates that they are numerical artefacts.
\item In the case $s=\frac p2$ the accuracy of the TN approximation gets worse with increasing $k$, while for $s\neq \frac p2$ it improves (for a smaller data set).
\end{itemize}

\begin{table}[!ht]
\centering
\resizebox{\textwidth}{!}{
\begin{tabular}{|c|c|c|c|c|c|c|c|c|c|c|c|c|}
\hline
$k$ & $\lambda_{1}$ & rel error & $\lambda_{2}$& rel error & $\lambda_{3}$ & rel error & $\lambda_{4}$& rel error & $\lambda_{5}$& rel error & $\lambda_{6}$ & rel  error  \\
\hline
2 & 0.5099813  &7.76$\%$ & 0.7115164 & 4.55$\%$ & 0.8183046 &3.2$\%$  & 0.8776668 &2.48$\%$  &  0.9129560 & 2.02$\%$ & 0.9352771 & 1.71$\%$ \\
&0.73610742& 100.09 $\%$&0.8307225& 78.18$\%$& 0.8844950& 62.34$\%$& 0.9170277&51.09 $\%$& 0.9378735& 42.94$\%$&0.9519037& 36.87 $\%$ \\
&2.05249738&118.20 $\%$&2.12030736& 92.76$\%$&2.15946205&74.40 $\%$&2.18366672& 61.30$\%$&2.19950490& 51.76$\%$&2.21036787&44.60 $\%$\\ 
&1.8843085 & 17.84$\%$&2.0334290& 15.44$\%$&2.1103984&13.11 $\%$&2.1537739& 11.19$\%$&2.1801354&9.68 $\%$&2.1971772& 8.49$\%$ \\
\hline
3 & 0.1644953 & 15.42$\%$ &  0.2071437 & 9.39$\%$  & 0.2254277 & 6.66$\%$ &  0.2343488 & 5.15 $\%$&  0.2392419 & 4.19 $\%$& 0.2421806 &3.53$\%$  \\
& 0.2131688 &168 $\%$&0.2282401& 115.4 $\%$&0.2358353& 85$\%$& 0.2401095& 66.4 $\%$&0.2427268& 54 $\%$& 0.2444367& 45.5 $\%$ \\
& 0.9917801 & 107.70$\%$&0.9948492& 76.86$\%$&0.9964864& 59.12$\%$&0.9974558& 47.82$\%$&0.9980752& 40.06$\%$&0.9985059& 35.50$\%$\\
&0.9835014 & 3.48$\%$&0.9911283&2.68 $\%$&0.9945269&2.15 $\%$&0.9963053&1.79 $\%$&0.9973446&1.52 $\%$&0.9980021&1.33 $\%$\\

\hline
4 & 0.0806650 &21.76$\%$  &  0.0969355 & 13.41$\%$ & 0.1032707 & 9.54$\%$ &  0.1062113 & 7.37$\%$ &  0.1077803 & 5.99$\%$  & 0.1087071 & 5.05$\%$ \\
& 0.0992625&212.9 $\%$  & 0.1043036& 136.2$\%$  &0.1067431&96.6 $\%$  &0.1080860& 73.9$\%$  &0.1088977 & 59.5$\%$  &0.1094250& 49.8$\%$  \\
\hline
5 & 0.0477721 &26.79$\%$ &  0.0559210 &16.49$\%$  & 0.0589276 & 11.71$\%$ &  0.0602881 & 9.02 $\%$&  0.0610043 & 7.32$\%$  & 0.0614286 &  6.61$\%$\\
& 0.0570821& 244.6$\%$ & 0.0594288& 149.5 $\%$ & 0.0605425& 103.8 $\%$ & 0.0611494& 78.5$\%$ & 0.0615242& 64.5$\%$ & 0.0618603& 78.5 $\%$ \\
\hline
6 & 0.0315512 & 30.73$\%$ &  0.0363237  & 18.82$\%$ & 0.0380257 & 13.30$\%$ & 0.0387841 & 10.22$\%$ &  0.0391840 & 8.77$\%$  & 0.0394826 & 21.67$\%$ \\
& 0.0369997& 268.1$\%$  &0.0383124& 158.8$\%$  &0.0389285& 108.8 $\%$  &0.0392705& 83.7$\%$  &0.0395636&  103.4$\%$  &0.0399422& 988.6$\%$  \\
\hline
7 & 0.0223782 & 33.93$\%$ &  0.0254703 & 20.66$\%$  & 0.0265475 & 14.55 $\%$&  0.0270234 & 11.25$\%$ &  0.0273029 & 16.65$\%$  & 0.0276131 & 138.03$\%$ \\
& 0.0259057& 286.3$\%$  &0.0267301& 165.8 $\%$  &0.0271160& 113.0 $\%$  &0.0273732& 107.4$\%$  &0.0276875& 513.4$\%$  & -& - \\
\hline
8 & 0.0166916 & 36.56$\%$ &  0.0188406 & 22.15$\%$  & 0.0195767 & 15.55 $\%$&  0.0199055 & 13.32$\%$ &  0.0201593 & 50.66$\%$  & - & - \\
& 0.0191417& 300.7$\%$  &0.0197019&171.1 $\%$  &0.0199741& 121.3 $\%$  &0.0202229& 207.4$\%$  & - & -& - & -  \\
\hline
9 & 0.0129257 & 38.75$\%$ &  0.0144977 & 23.36 $\%$ & 0.0150295 & 16.40 $\%$&  0.0152837 & 19.98$\%$ &  0.0155447 & 234.83 $\%$ & - & - \\
& 0.0147169&312.5 $\%$ &0.0151206& 175.7$\%$ &0.0153413& 144.3$\%$ &0.0156005& 1570.4 $\%$ &- & - &- & - \\
\hline
10 & 0.0103028 & 40.60$\%$ &  0.0114992 & 24.37 $\%$ & 0.0119010 & 17.37$\%$ &  0.0121221 & 37.56$\%$ &  - &  -& - & - \\
& 0.0116655& 322.2$\%$ &0.0119712& 181.1$\%$ &0.0121731&202.3 $\%$ & - & - & - & - &- &- \\
\hline
\end{tabular}}
\caption{The numerically computed first six eigenvalues  for $N=1$ and a range of values for the parameter $k$, for the four cases $(s,j,p)=(1,1,2),(1,2,2),(0,2,2),(0,0,2)$ from top to bottom in each cell. For the latter two cases, bound states only occur for the first two values of $k$ according to the condition \eqref{boundcond}, illustrated in Figure~\ref{fig:sintp2}. The relative errors of the TN approximation  \ref{relerror} are shown as  percentages.}
\label{tab:data12}
\end{table}

\section{Conclusions} 

In this paper we generalised  and extended the study of spectral properties of the  gauged TN  and  ES geometries  initiated in \cite{JS2} and \cite{JS3}, to a parameter-dependant family of gauged TB geometries which interpolates between them and combined various non-trivial features of both. We showed that the gauged Laplace operators of all members of the family have qualitatively similar spectra, including  an infinite tower of Coulombic bound states  when the fibre angular momentum $s$  satisfies a bound set by the Chern class of the  connection. Our findings are underpinned by a rigorous analysis of selfadjoint extensions which  
also applies to the  previous studies of  the radial TN and ES Laplacians.  Remarkably, the edge-cone singularities in the classical geometry of the TB family cause no particular difficulty for  the  spectral analysis.

One surprising aspect of our study is the accuracy of the analytical approximation to the spectrum provided by the exactly solvable  gauged TN Laplacian, and its dependence on the total  angular momentum  $j$ and the fibre angular momentum $s$. The gauged TN Laplacian has a dynamical symmetry of  Runge-Lenz type \cite{JS2}, and as a result its eigenvalues do not directly depend on the value of the total angular momentum. This symmetry is not present in the TB family:  its spectrum shows no unexpected additional degeneracy and does depend on $j$. It is therefore not surprising that the TN approximation is better   for  some choices of $j$ than for others. However, it is not clear  why the choice $j=s$ is particularly advantageous, and this should be investigated in future research.

A further direction for future work is  the consideration of negative values of the parameter $N$ which was kept fixed and positive here. The TN metric for negative $N$ - the  `negative mass TN space'   - has a well-defined Laplacian even though the classical geometry has a singularity, and its spectrum can be determined analytically \cite{GM}.  Remarkably, this Laplacian has infinitely many Coulombic bound states even without the presence of a connection. It would therefore  be of interest  to consider the case of negative $N$ for all of the geometries discussed here,  to see if the Laplacian can be given a satisfactory definition and to study its spectrum. The interplay of the parameters $k$ and $N$  and their interaction with a self-dual connection on these spaces provide a rich field for further study. 

\vspace{1cm} 
\noindent {\bf  Acknowledgements.} Funding for the research reported in this paper was provided by EPSRC grants EP/K00848X/1 and EP/L504774/1. BJS thanks Guido Franchetti for useful discussions.

\end{document}